\documentclass[showpacs,aps,prb,twocolumn,superscriptaddress]{revtex4-1}

\usepackage{graphicx}
\usepackage{dcolumn}
\usepackage{bm}

\begin{document}



\title{Behavior of hydrogen ions, atoms, and molecules in $\alpha$-boron studied
using density functional calculations}

\author{Philipp Wagner}
 \email{philipp.wagner@cnrs-imn.fr}
\author{Christopher P. Ewels}
 \email{chris.ewels@cnrs-imn.fr}
 \homepage{http://www.ewels.info/}
\affiliation{Institut des Mat\'eriaux Jean Rouxel, Universit\'e de Nantes, CNRS UMR 6502, 44322 Nantes, France}
\author{Irene Suarez-Martinez}
\affiliation{Institut des Mat\'eriaux Jean Rouxel, Universit\'e de Nantes, CNRS UMR 6502, 44322 Nantes, France}
\affiliation{Nanochemistry Research Institute, Curtin University of Technology, Perth, WA 6845, Australia}
\author{Vincent Guiot}
\affiliation{Institut des Mat\'eriaux Jean Rouxel, Universit\'e de Nantes, CNRS UMR 6502, 44322 Nantes, France}
\author{Stephen F. J. Cox}
\affiliation{Condensed Matter and Materials Physics, University College London, WC1E 6BT, UK}
\affiliation{ISIS Facility, Rutherford Appleton Laboratory, Chilton, Oxfordshire OX11 0QX, UK}
\author{James S. Lord} 
\affiliation{ISIS Facility, Rutherford Appleton Laboratory, Chilton, Oxfordshire OX11 0QX, UK}
\author{Patrick R. Briddon}
\affiliation{School of Natural Sciences, University of Newcastle upon Tyne, Newcastle upon Tyne, UK}

\date{\today}

\begin{abstract}
\textbf{For full reference please see: Phys. Rev. B 83, 024101, 2011}\\
\\
We examine the behaviour of hydrogen ions, atoms and molecules in $\alpha$-boron using density functional calculations. 
Hydrogen behaves as a negative-U centre, with
positive H ions preferring to sit off-center on inter-layer bonds and negative H ions sitting preferably
at in-plane sites between three B$_{12}$ icosahedra.  
Hydrogen atoms inside B$_{12}$ icosahedral cages are unstable, drifting off-center 
and leaving the cage with only a 0.09~eV barrier.  
While H$^0$ is extremely mobile (diffusion barrier 0.25~eV), H$^+$ and H$^-$ have higher diffusion barriers of 0.9~eV.
Once mobile these defects will combine, forming H$_2$ in the interstitial void space, 
which will remain trapped in the lattice until high temperatures.
Based on these results we discuss potential differences for hydrogen behaviour in $\beta$-boron, 
and compare with experimental muon-implantation data.
\end{abstract}


\maketitle

Boron is a fascinating elemental material about which we are still making many new discoveries.  Only recently a new form of Boron was found, showing unusual negative-U behaviour between B$_{12}$ and B$_2$ sub-units within the crystal \cite{Oganov2009}. Sitting somewhere between metals and insulators on the periodic table, the bonding in boron is strongly dependent on local environment and factors such as temperature, pressure and impurities.  The simplest such impurity is hydrogen, yet to date no studies exist of the behaviour of hydrogen in bulk boron.

To date 16 allotropes of elemental boron have been reported. The best known crystal phases are the $\alpha$-rhombohedral structure ($\alpha$-boron) and the $\beta$-rhombohedral structure ($\beta$-boron), as well as two tetragonal modifications which are probably stabilized by foreign atoms \cite{Albert2009}. The $\alpha$-boron structure as the simplest of them contains 12 atoms in a rhombohedral unit cell forming a slightly distorted icosahedron (B$_{12}$, see Figure \ref{Hinalphapic}). More complex, but thermodynamically stable at high temperatures is the $\beta$-boron crystal with a unit cell containing 105 atoms \cite{Shang2007}. Compared with the $\alpha$-boron phase, $\beta$-boron bulk is less dense and softer \cite{Masago2006}. An even more stable modification of $\beta$-boron has been proposed with 106 atoms in the unit cell \cite{Setten2007}.
Boron based nanostructures have also been proposed \cite{Boustani1999}, and boron single and double wall nanotubes and boron nanoribbons have been reported \cite{Ciuparu2004,Sebetci2008,Xu2004}.

Boron compounds have been proposed as promising candidates for hydrogen-storage \cite{Berg2008}, for example alkali doped boron sheets as possible hydrogen acceptors \cite{Er2009}. Although such applied studies are already underway, there still remains much to understand concerning hydrogen behaviour in the crystalline boron phases, which forms the subject of the current article. Given that the B$_{12}$-icosahedron is the basic element for both rhombohedral modifications, we have focused on the simpler $\alpha$-boron phase.  Based on these results comparable sites for the $\beta$-boron phase are discussed.

\section{Computational Details}

We perform density functional theory calculations under the local density approximation as implemented in the AIMPRO code \cite{AIMPRO, AIMPRO2}. The calculations were carried out using supercells, fitting the charge density to plane waves within an energy cut-off of 300 Ha. Electronic level occupation was obtained using a Fermi occupation function with $kT$ = 0.04~eV. Relativistic pseudo-potentials are generated using the Hartwingster-Goedecker-Hutter scheme \cite{HGH}. These functions are labelled by multiple orbital symbols, where each symbol represents a Gaussian function multiplied by polynomial functions including all angular momenta up to maxima $p$ ($l$ = 0, 1) and $d$ ($l$ = 0, 1, 2). Following this nomenclature, the basis sets used for each atom type were $pdpp$ (B) and $ppp$ (H) (a more detailed account of the basis functions can be found elsewhere \cite{Goss2007}). A Bloch sum of these functions is performed over the lattice vectors to satisfy the periodic boundary conditions of the supercell. Structures were geometrically optimized with a single k-point for the B$_{12}$ cluster in a large 23.81 {\AA} cubic supercell, and a 4$\times$4$\times$4 k-point grid for pure $\alpha$-boron. Both rhombohedral and triclinic lattices were examined to find the optimal structure. Hydrogen in boron was modelled using a 2$\times$2$\times$2 k-point grid for a triclinic cell containing 8 B$_{12}$ icosahedra, {\it i.e.} B$_{96}$H$_n$, $n=1,2$.
H$^+$ and H$^-$ were modelled in charge neutral unit cells by the inclusion of a uniform countering background charge. Diffusion barriers were determined using the nudged elastic band method \cite{Henkelman2000}.  Saddle points were checked via additional energy double derivative calculations giving a single negative frequency.

The majority of the energies considered below are relative, for example comparison of different stable sites for H$^+$ within the lattice, or calculated migration barriers.
There are a number of different ways to correct absolute binding energies to take into account interaction between charged defects in neighbouring unit cells, and the literature is far from being clear on the best approach.  While the standard reference for some time has been \citet{Makov1995}, more recently other approaches are under development\cite{Freysoldt2009,Lany2009}.  
The base interaction energy can be estimated from the Madelung energy for an array of point charges with a neutralizing background as $E = - \alpha q^2 / (2 \epsilon L)$, and a reasonable approximation to the charged cell correction is then given in \cite{Lany2009} as $\approx 2 E / 3$.  
While corrections can be reasonably large for defects with higher charge states, as our H ions are charge q=$\pm$1 and E varies with q$^2$ these corrections are small.  Our repeated unit cell sizes used for the H insertion calculations are also large giving 
$L = \Omega^{1/3} = 8.74\AA$.  Taking the $\beta$-boron dielectric constant $\epsilon \approx$10 and an overestimate for the Madelung constant $\alpha$ of 10 only results in corrections to formation energies of $\approx$0.04eV.  This correction is much smaller than any of the energy differences we calculated between different hydrogen arrangements.  

\section{Hydrogen addition to isolated B$_{12}$ icosahedron}

\begin{figure}
\includegraphics[width=8cm]{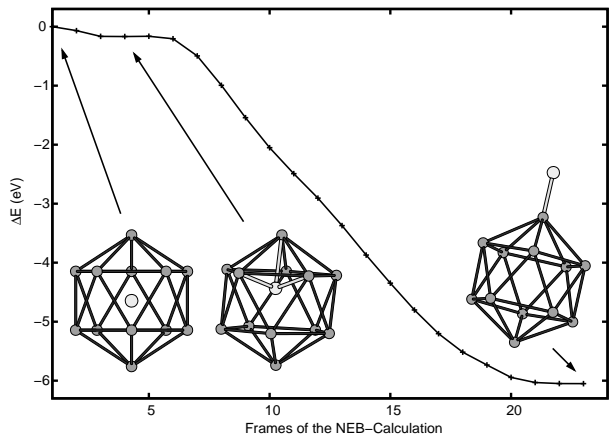}
\caption{Diffusion barrier (eV) for H to leave an isolated B$_{12}$ cluster. \label{neb}}
\end{figure}

As a basic element of the $\alpha$-boron structure we first examine an isolated B$_{12}$ cluster. Without hydrogen it forms a perfect icosahedron with B-B bondlengths of $d_{\mathrm{B}_{12}}=1.68$ \AA, in agreement with literature values (1.7 {\AA}\cite{Caputo2009}, 1.71 {\AA}\cite{Fujimori1997}). Adding hydrogen to the B$_{12}$, the energetically most stable position is with hydrogen bound to the outside of B$_{12}$ ($d_{\mathrm{B-H}}=1.19$ {\AA}), deforming the icosahedron giving bondlengths in the range $d_{\mathrm{B}_{12}}=1.62-1.74$ {\AA}. An earlier study found hydrogen stable in the center of the B$_{12}$ cluster \cite{Hayami1999}, however our energy calculations (Figure \ref{neb}) show that this site is actually a metastable maximum (with a slightly expanded B$_{12}$ to $d_{\mathrm{B}_{12}}=1.72-1.73$ {\AA}). Any off-site motion moves the H to an interior site near a B-B-B triangle of the B$_{12}$ ($d_{\mathrm{B}_{12}}=1.69-1.83$ {\AA}). The energy difference between these two sites is only 0.17 eV, while the position outside the B$_{12}$ cluster is 6.05~eV more stable than the center position.  Hydrogen inside the B$_{12}$ cluster migrating to the most stable configuration only has to overcome a barrier energy of $\Delta E_B = 0.0042$~eV. This small value indicates that hydrogen atoms will sit covalently bonded to the outside of isolated B$_{12}$ clusters.

\section{Single hydrogen atoms in $\alpha$-boron}

\begin{figure}[b]
\begin{flushleft}
(a)
\end{flushleft}
\vspace{-0.2cm}
\includegraphics[width=7cm]{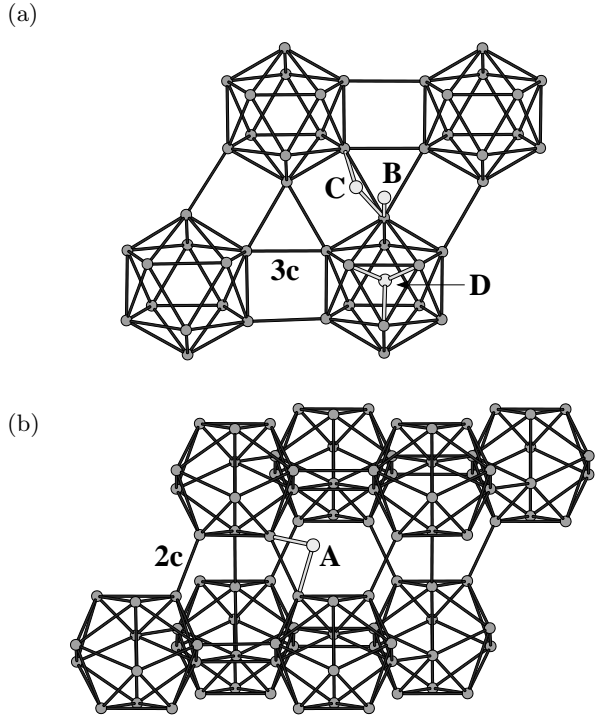}\\
\vspace{-4.5cm}
\begin{flushleft}
(b)
\end{flushleft}
\vspace{3.5cm}
\caption{(a) stable sites B-D of H$^0$ (shown white) in the $\alpha$-boron structure with three-center bonds (3c), (b) site A near a two-center bond (2c) shown in a sideview of the $\alpha$-boron structure. \label{Hinalphapic}}
\end{figure}

We next consider $\alpha$-boron. For the following discussion we follow the nomenclature to describe inter-cage bonding adopted by \cite{Vast1997,Oganov2009,Fujimori1999}, {\it i.e.} inter-layer bonds are referred to as 2-center (2c) bonds, intra-layer bonds as 3-center (3c) bonds (see Figure \ref{Hinalphapic}).   It should not be confused with the nomenclature developed for boron-hydrogen molecular compounds by Lipscomb {\it et al} \cite{Dixon1977,Brown1977,LonguetHiggins1949},  notably their chemical two- and three-center bonds refer to the number of boron atoms involved in chemical bonding, rather than the topology of the boron cage lattice as in our case.

Using a rhombohedral supercell we calculated lattice parameters to be $\alpha= 58.15^\circ$ and $a= 4.98$ {\AA} and B-B bond lengths for the B$_{12}$ icosahedra of $d_{\mathrm{B}_{12}}=1.72-1.78$ {\AA} with inter-icosahedral two-center bonds (2c) $d_{2c}=1.65$ {\AA} and three-center bonds (3c) $d_{3c}=1.98$ {\AA}. The resulting optimised structure is in very good agreement with calculations from \citet{Vast1997} and very similar to other cases \cite{Kawai1991,Masago2006}. Bond lengths using a very similar triclinic supercell ($a=4.83$ {\AA}, $b=4.83$ {\AA}, $c= 4.98$ {\AA}, $\alpha= 60.12^\circ$, $\beta=60.95^\circ$, $\gamma=60.97^\circ$), also with a $4 \times 4 \times 4$ k-point mesh found exactly the same inter-icosahedral 2c- and 3c-bonds lengths and $d_{\mathrm{B}_{12}}=1.72-1.77$ {\AA} for the B$_{12}$ icosahedra. The volume of the triclinic cell is 0.09 \% smaller than the rhombohedral, and the total energy of the cell is 0.0106~eV lower. We therefore used the triclinic cell for all subsequent implantation calculations.

A single neutral H atom was then added at different sites and the structures optimised again, this time with fixed lattice parameters. 
For H$^0$ four stable sites (A-D) were found, as shown in Figure \ref{Hinalphapic}a,b. In site A the H atom is located near a 2c-bond with bond lengths $d_{\mathrm{H_A-B}}=1.35$ {\AA} and $d_{\mathrm{H_A-B}}=1.26$ {\AA}. In site B the H atom sits centered over a boron triangle of 3c-bonds between the icosahedra ($d_{\mathrm{H_B-B}}=1.27$ {\AA}). Site C is near site B, but the H atom lies directly over a 3c-bond between two icosahedra ($d_{\mathrm{H_C-B}}=1.41$ {\AA} for both). 
As for isolated B$_{12}$ we also found a much less stable metastable site within the B$_{12}$ cage (see Table \ref{Hinalpha}), position D ($d_{\mathrm{H_D-B}}=1.28$ {\AA} for all three), where the hydrogen lies near a boron triangle pointing towards a 2c-bond.  When placed in other sites within the B$_{12}$ icosahedron (including the center site) the hydrogen moves to more stable locations outside the icosahedron without any energy barrier. The energy barriers to move between site D and either site A or B (Table \ref{barr}) are very small (0.13~eV and 0.09~eV respectively), and thus as for the isolated B$_{12}$, H will not remain within the cage except possibly at liquid helium temperatures.  This is in agreement with recent calculations for much higher hydrogen densities (1 per B$_{12}$) \cite{Caputo2009}, although we find a different final stable location outside the cage.
As well as the space at icosahedra centers the $\alpha$-boron lattice also has an interstitial void space between the B$_{12}$ layers (see Figure \ref{Hinalphapic}), however once again hydrogen is expected to be metastable here and will displace to more stable neighbouring bond sites.

\begin{table}
 \caption{Sites for H$^0$, H$^+$ and H$^-$ in $\alpha$-boron and their calculated stability $\Delta$E (eV) relative to the most stable site at that charge state (as shown in Fig. \ref{Hinalphapic}). \label{Hinalpha}}
 \begin{ruledtabular}
 \begin{tabular}{l|c c c}
 Site & H$^0$ $\Delta$E (eV) & H$^+$ $\Delta$E (eV)& H$^-$ $\Delta$E (eV)\\
 \hline
 $A$/$A_+$/$A_-$& 0.0 & 0.0 & +0.95 \\
 $B$/$B_+$/$B_-$ & +0.17 & +0.48 & 0.0\\
 $C$/-/- & +0.24 &  - & - \\
 $D$/$D_+$/$D_-$ & +2.35 & +2.31 & +3.39\\
 -/-/$B*$ &   -     &    -    & +0.56 \\
 Inter. Void & +0.60 & +1.82 & +0.73 \\
 \end{tabular}
 - : site is not stable in this configuration
 \end{ruledtabular}
 \end{table}
\begin{figure}
(a) \hspace{4.5cm} (b)\\ 
\includegraphics[width=8cm]{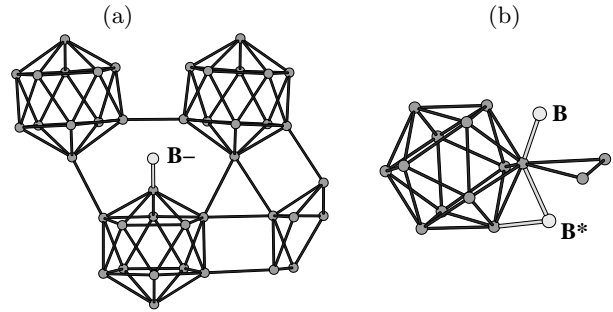}
\caption{The most stable sites for (a) H$^-$ (site $B_-$) and (b) H$^0$ (site $B$) compared with site $B*$ for H$^-$ over and under a 3c bond triangle. \label{H-inalphapic}}
\end{figure}

We next determine the stable sites for H$^+$ and H$^-$ (sites marked as $A_x$, $B_x$,.. with $x=+,-$). For H$^+$ stable sites were found to be comparable to that of neutral hydrogen, with the most stable location near a 2c-bond (site $A_+$, $d_{\mathrm{H_{A+}-B}}=1.27$ {\AA} for both). Site $B_+$  is also stable ($d_{\mathrm{H_{B+}-B}}=1.46$ {\AA} for all three bonds, directly centered between the 3c bond triangle B atoms and inducing an enlargement of the 3c-bonds underneath to $d_{3c}=2.26$ {\AA}), but with a relative energy difference of +0.48~eV to site $A_+$.  For H$^+$ site C is not stable (see Table \ref{Hinalpha}).
H$^-$ behaves differently and this time the most stable site is $B_-$ (see Table \ref{Hinalpha}). As for H$^+$ the hydrogen sits between the icosahedra and induces a separation of the icosahedra with $d_{3c}=2.28$ {\AA} (see Figure \ref{H-inalphapic}a). However H$^-$ forms a shorter bond ($d_{\mathrm{H_{B-}-B}}=1.2$ {\AA}) and sits off-centered over the triangle space, pointing more into the interstitial void.
Unlike H$^0$ and H$^+$ there is also a new site called $B*$ near the B$_{12}$ icosahedron on the other side of the triangle. On this side of the triangle the H$^-$ lies over a boron bond of the B$_{12}$-icosahedron with bond lengths $d_{\mathrm{H_{B*}-B}}=1.35$ {\AA} and $d_{\mathrm{H_{B*}-B}}=1.50$ {\AA} (the difference between site $B*$ and site B for H$^0$ is shown in Figure \ref{H-inalphapic}b). Site $A_-$ is a metastable position switching ether to $B_-$ or $B*$. Site C was not found to be stable for H$^-$.
As for H$^0$, site $D_+$ and $D_-$ (the off-centered site within the icosahedron) are energetically unfavourable for both H$^+$ and H$^-$, and we suppose hydrogen leaves the cage to sit in position $A_+$ or respectively $B_-$.
 
Thus in summary, the most stable site for H$^+$ and H$^0$ in $\alpha$-boron is just off the 2c-bond between layers of B$_{12}$ icosahedra, while H$^-$ prefers to sits in-plane between three icosahedra, slightly separating them.\\

\begin{table}
 \caption{Calculated diffusion barriers $\Delta E_B$ (eV) for H$^0$, H$^+$ and H$^-$ in $\alpha$-boron. (Path indicates the barrier which has to overcome passing from site $x$ to site $y$: $x \rightarrow y$) \label{barr}}
 \begin{ruledtabular}
 \begin{tabular}{c c | c c | c c}
  \multicolumn{2}{c|}{H$^0$} &  \multicolumn{2}{c|}{H$^+$} &  \multicolumn{2}{c}{H$^-$} \\
  \hline
  Path &  $\Delta E_B$& Path &  $\Delta E_B$& Path & $\Delta E_B$\\
  $D\rightarrow A$& 0.13 &$B_+ \rightarrow A_+$& 0.44 &$A_- \rightarrow B_-$ & 0.03\\
  $D\rightarrow B$& 0.09 & & &$B* \rightarrow B_-$& 0.39 \\
  $B\rightarrow A$  & 0.08 & & & &\\
  \end{tabular}
 \end{ruledtabular}
 \end{table}

We next calculate diffusion barriers between the most stable sites for isolated H$^0$, H$^+$ and H$^-$ (see Table \ref{barr}).  
H$^0$ has a very low diffusion barrier of 0.25~eV ($0.17+0.08$) and hence will be highly mobile in the inter-layer region of $\alpha$-boron. H$^+$ also migrates between interlayer sites but with a diffusion barrier of 0.92~eV ($0.48+0.44$).  H$^-$ however migrates within the layers between 3c-bond sites, also with a barrier of 0.95~eV ($0.56+0.39$). 

\section{Hydrogen pairs and H$_2$}

Comparison of the total energies for two isolated H atoms in boron shows that the most stable configuration is not with two H$^0$ but rather one H$^+$ and one H$^-$, {\it i.e.} H in $\alpha$-boron is a {\it negative U} center (see Table \ref{tab:H2}).  This suggests that low density atomic H will adopt an overall charge neutral mixture of H$^+$ and H$^-$ centers, with no net doping of the lattice. While neutral H would be highly mobile within the lattice, the thermodynamically preferred coexistence of H$^+$ and H$^-$ centers should only just be mobile at room temperature.  We note that, depending on the defect density and associated degree of electronic coupling between them via the host boron lattice, it may be possible to photoexcite H$^+$/H$^-$ into a metastable H$^0$/H$^0$ state, as has been shown recently for N$_{2}$H centers in diamond \cite{Goss2010}.

While doping $\alpha$-boron could encourage charge compensation by interstitial H and result in a concentration bias of either H$+$ or H$^-$, in preliminary test calculations we found neither potassium or vanadium were stable within the B$_{12}$ icosahedra.

\begin{figure}
(a) \hspace{4.5cm} (b)\\ 
\includegraphics[width=8.5cm]{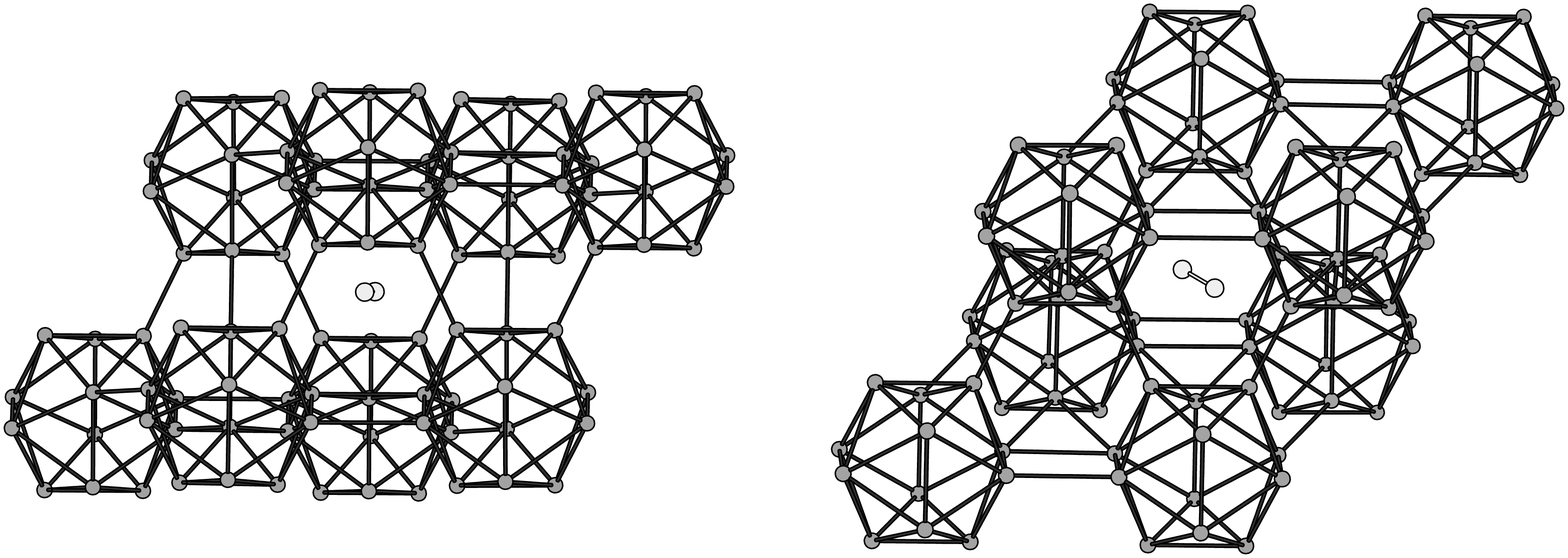}
\caption{$H_2$ in the hexagonal interstitial site between icosahedral layers, (a) side and (b) top view. \label{H2pic}}
\end{figure}

H$^+$/H$^-$ centers will experience a Coulombic attraction and we might expect H$^+$ and H$^-$ to migrate towards one another. Table~\ref{tab:H2} shows that placing two H atoms in the same inter-icosahedral triangle (not dissimilar to H$_2^*$ in silicon \cite{Jackson1991}) lowers the net system energy by 0.84~eV compared to an infinitely separated H$^+$/H$^-$ pair.  These neighbouring pairs can then combine to create an H$_2$ molecule, releasing a further 0.79~eV, which sits in the void space between icosahedral layers (see Figure \ref{H2pic}).  However it is highly constrained, and the system could lower its energy by a further 1.70~eV if H$_2$ was able to migrate to the boron surface and escape.  

The $\alpha$-boron interlayer space contains two types of interlayer void.  The first and largest lies between 3c bond triangle sites in neighbouring layers, where the H$_2$ sits (see Figure \ref{H2pic}).  
These alternate with a smaller space between a 3c bond triangle in one layer and the base of an icosahedron in the other.  Any interlayer diffusion path must necessarily pass through both sites.
Our calculations show that H$_2$ is unable to migrate as a molecule, instead dissociating into a H$^+$/H$^-$ pair which then diffuse through this smaller void before recombining.  The NEB calculations give a total barrier of $\approx$2.2~eV, consisting of an $\approx$1.4~eV dissociation energy followed by $\approx$0.9~eV migration barriers for each hydrogen.  This means H$_2$ will be trapped and immobile in the large voids until high temperatures.  We note that we did not find any other stable sites for H$_2$ in the crystal, suggesting a maximum possible H$_2$ concentration in $\alpha$-boron of B$_{24}$H$_2$.
The initial H$_2$ separation into H$^+$/H$^-$ is barrierless, showing that H$_2$ formation from H ions on neighbouring sites can also occur spontaneously.
We note that we know of no other material where H$_2$ is stable yet migrates by dissociation.  The behaviour lies between that of diamond, where the H$_2$ molecule is not stable, and silicon, where H$_2$ is stable and can migrate as a molecule.  

\begin{table}
\caption{\label{tab:H2}Comparison of stability for different arrangements of hydrogen
pairs within $\alpha$-boron. H$^x$ refers to hydrogen in its most stable location
with this charge state ({\it i.e.} inter-cage sites).}
\begin{ruledtabular}
\begin{tabular}{l c}
Hydrogen Arrangement & $\Delta E$ (eV)\\
\hline
H$_2$ molecule separated from boron & -\\
H$_2$ inside the boron lattice & +1.70\\
H$^+$/H$^-$ ``intimate pair'' (same triangle) & +2.49\\
H$^+$/H$^-$ infinitely separated in the lattice & +3.33\\
H$^0$/H$^0$ infinitely separated in the lattice & +4.12\\
\end{tabular}
\end{ruledtabular}
\end{table}

\section{Discussion and Conclusions}

Our results for the $\alpha$-phase allow us to speculate on hydrogen behaviour in the more
complex $\beta$-boron phase.  $\beta$-boron has no equivalents of the $\alpha$-boron 3c bond triangles,
however bonds with similar lengths to the 2c- and 3c- bonds exist.  Thus we would
expect no equivalent of the B, B$^*$ and B$^-$ sites for hydrogen, and the most stable sites are likely instead
to be similar to A (2c-bonds) or C (3c-bonds).  This agrees with results from implantation of positive muonium into
$\beta$-boron \cite{Cox2009}.   We can also suppose that H will
not sit within the B$_{12}$ icosahedra in the $\beta$-boron phase.  Unlike $\alpha$-boron there are also some
under-coordinated partial vacancy sites likely to trap hydrogen, and given the larger connected void
spaces we assume H$_2$ migration will have a significantly lower barrier.\\

Table \ref{tab:H2} shows that hydrogen within the $\alpha-$boron lattice is less stable than isolated gas phase H$_2$.  
Thus hydrogen incorporation within the lattice will only occur through kinetic restrictions, {\i.e.} if hydrogen becomes trapped in the lattice,
either during growth or through implanation, and is then not able to escape.   Since $\beta-$boron has larger void spaces we assume that H$_2$
will be more stable in the $\beta-$boron void sites.  Thus the presence of significant quantities of hydrogen within the $\alpha-$boron lattice
may serve to destabilise it with respect to the less dense $\beta-$boron phase.
This may be a contributing factor to the observed difficulty in synthesising $\alpha-$boron, as well as the $\alpha\rightarrow\beta-$boron phase transition 
(we note that molecular hydrogen in other crystalline semiconductors such as silicon is often difficult to identify).  

Thus we have now a first picture of hydrogen behaviour in pristine $\alpha$-boron.  
At low densities H atoms will form a reservoir of H$^+$ and H$^-$ ions in the inter-icosahedral space, resulting in a charge neutral crystal.  
Once these become mobile, either through thermal activation or via excitation into highly mobile metastable H$^0$, they will diffuse together, creating interstitial molecular H$_2$.  This will remain trapped in interstitial void spaces until at very high temperatures when it will diffuse out.  These results suggests that hydrogen doping of boron will be difficult and probably only possible through ion implantation rather than gas phase impregnation and thus $\alpha$-boron is unlikely to serve as a useful hydrogen storage material.  
There may also be the possibility of H diffusion and resultant H$_2$ formation at very low temperatures due to quantum tunnelling.

\begin{acknowledgments}
PhW and CPE thank the NANOSIM\_ GRAPHENE project n$^{\circ}$ANR-09-NANO-016-01 funded by the French National Agency (ANR) in the frame of its 2009 programme in Nanosciences, Nanotechnologies \& Nanosystems (P3N2009).  We thank the CCIPL where some of these calculations were performed.
\end{acknowledgments}

\bibliographystyle{apsrev4-1}

%

\end{document}